\documentclass[aps,prl,twocolumn]{revtex4-1}

\usepackage{ifthen}

\usepackage{amssymb}
\usepackage{color}
\usepackage{graphics}
\usepackage{hyperref}

\newcommand{\psec}[1]{\emph{#1.}---}

\newcommand{\bq}{\begin{equation}}
\newcommand{\eq}{\end{equation}}
\newcommand{\bqa}{\begin{eqnarray}}
\newcommand{\eqa}{\end{eqnarray}}

\newcommand{\scal}{\varphi}

\newcommand{\Geff}{G_{\rm eff}}
\newcommand{\Msun}{M_\odot}
\newcommand{\rhom}{\rho_{\rm m}}
\newcommand{\drhom}{\delta\rhom}
\newcommand{\bscal}{\bar{\scal}}
\newcommand{\scalout}{\varphi_{\rm out}}
\newcommand{\scalin}{\varphi_{\rm in}}
\newcommand{\scalenv}{\varphi_{\rm env}}

\newcommand{\Deltavir}{\Delta_{\rm vir}}
\newcommand{\rhos}{\rho_{\rm s}}
\newcommand{\rs}{r_{\rm s}}
\newcommand{\rc}{r_{\rm c}}

\begin{document}

\title{How chameleons core dwarfs with cusps}

\author{Lucas~Lombriser}
\affiliation{Institute for Astronomy, University of Edinburgh, Royal Observatory, Blackford Hill, Edinburgh, EH9~3HJ, U.K.}
\author{Jorge~Pe\~{n}arrubia}
\affiliation{Institute for Astronomy, University of Edinburgh, Royal Observatory, Blackford Hill, Edinburgh, EH9~3HJ, U.K.}

\date{\today}

\begin{abstract}

The presence of a scalar field that couples nonminimally and universally to matter can enhance gravitational forces on cosmological scales while restoring general relativity in the Solar neighborhood.
In the intermediate regime, kinematically inferred masses experience an additional radial dependence with respect to the underlying distribution of matter, which is caused by the increment of gravitational forces with increasing distance from the Milky Way center.
The same effect can influence the internal kinematics of subhalos and cause cuspy matter distributions to appear core-like.
Specializing to the chameleon model as a worked example, we demonstrate this effect by tracing the scalar field from the outskirts of the Milky Way halo to its interior, simultaneously fitting observed velocity dispersions of chemo-dynamically discriminated red giant populations in the Fornax and Sculptor dwarf spheroidals.
Whereas in standard gravity these observations suggest that the matter distribution of the dwarfs is cored, we find that in the presence of a chameleon field the assumption of a cuspy Navarro-Frenk-White profile becomes perfectly compatible with the data.
Importantly, chameleon models also predict the existence of slopes between two stellar subcomponents that in Newtonian gravity would be interpreted as a depletion of matter in the dwarf center.
Hence, an observation of such an apparently pathological scenario may serve as a smoking gun for the presence of a chameleon field or a similar modification of gravity, independent of baryonic feedback effects.
In general, measuring the dynamic mass profiles of the Milky Way dwarfs provides stronger constraints than those inferred from the screening scale of the Solar System since these are located at greater distances from the halo center.

\end{abstract}

%%%%%%%%%%%%%%%%%%%%%%%%%%%%%%%%%%%%%%%%%%%%%%%%%%%%%%%%%%%%%%%%%%%%%%%%%%%%%

\maketitle

%%%%%%%%%%%%%%%%%%%%%%%%%%%%%%%%%%%%%%%%%%%%%%%%%%%%%%%%%%%%%%%%%%%%%%%%%%%%%

\psec{Introduction}
The observed late-time acceleration of the expansion of our Universe constitutes one of the greatest puzzles to science.
Cosmic acceleration may be driven by a cosmological constant introduced by vacuum energy.
Its observed value is, however, theoretically not understood, being $\mathcal{O}(10^{60}-10^{120})$ smaller than expected.
Alternatively, the effect may be caused by a dark energy in form of a scalar field.
The presence of a scalar field can be motivated as an effective representation of models that attempt to unify general relativity with the Standard Model interactions.
If this scalar field is nonminimally coupled, it introduces a modification of gravity on large scales which, however, must be shielded in the Solar System where general relativity is well tested~\cite{will:05}.
Therefore, viable modifications of gravity generically predict a transition regime, which interpolates between modified forces acting on large scales, and the local region in which modifications are shielded and general relativity is recovered.

Besides cosmic acceleration, another well known problem in astronomy is the observation of low-mass halos indicating cored matter distributions whereas collisionless $N$-body simulations of cold dark matter predict a universal, cuspy profile.
Typically, this discrepancy is attributed to baryonic feedback processes, which can turn a cuspy into a cored matter density profile~\cite{navarro:96,mashenko:07,pontzen:11,zolotov:12,teyssier:12,dicintio:13,pontzen:14}.
However, in low-luminosity, dark matter dominated galaxies in which star formation has been inefficient the impact of feedback cannot be strong~\cite{penarrubia:12,weinberg:13,madau:14}.
This may be the case in dwarf spheroidals, for which cores have been inferred~\cite{walker:11,amorisco:11} (see, however~\cite{strigari:14}), possibly posing a problem to baryonic explanations.

Here, we show that the appearance of core-like dynamic mass profiles can be an expected consequence of exposure to enhanced gravitational forces experiencing a shielding mechanism.
To demonstrate this point, we specialize our discussion to the well studied chameleon models~\cite{khoury:03a}, which may be viewed as representative models exhibiting a transition effect~\cite{vainshtein:72,babichev:09,hinterbichler:10,lombriser:14b} generic to viable large-scale modifications of gravity, although we note that the requirement of local compatibility excludes the chameleon field as cause of the late-time acceleration~\cite{wang:12}, so that a cosmological constant or dark energy needs to be reintroduced.
Assuming a universal, cuspy Navarro-Frenk-White (NFW)~\cite{navarro:95} dark matter density profile, we then trace the chameleon field from the Local Group into the Milky Way halo and its subhalos, simultaneously reproducing the velocity dispersions of chemo-dynamically distinct red giant populations in the Fornax and Sculptor dwarf spheroidals.

\psec{Chameleon models}
In addition to the metric tensor field, in chameleon gravity~\cite{khoury:03a} we consider a scalar field that couples nonminimally and universally to all matter species.
Gravitational forces then become enhanced due to the contribution of its gradient.
However, the effective scalar field potential in chameleon gravity is such that in its minimum the scalar field scales inversely proportional to curvature.
Hence, in high-curvature regions gravitational modifications get suppressed and general relativity is restored.
Similarly, gravity returns to standard at low curvature and scales larger than the Compton wavelength associated with the scalar field at the cosmological background.
In between, the chameleon field enhances gravitational forces and modifies the growth of structure and its internal kinematics.

Here, we are interested in the chameleon force within a spherically symmetric matter distribution of the form
\bq
  \drhom(r) = \frac{\rhos}{(r/\rs)^{\gamma}(1+r/\rs)^{3-\gamma}}, \label{eq:density}
\eq
where $\rhos$ and $\rs$ is the characteristic density and radius, respectively.
For $\gamma=0$, the interior of $\drhom(r)$ is cored whereas it is cusped for $\gamma>0$.
The case $\gamma=1$ corresponds to the NFW profile, which has also been shown to provide good fits to collisionless $N$-body simulations of cold dark matter in chameleon gravity~\cite{lombriser:12}.
The chameleon field profile within this matter distribution can be derived from the scalar field equation, which follows from variation of the chameleon scalar-tensor action with respect to the scalar field.
In the quasistatic limit it can be approximated by the outer and inner solution~\cite{pourhasan:11,lombriser:12}
\bq
 \scal = \left\{ \begin{array}{ll} \scalout, & r>\rc, \\ \scalin, & r\leq\rc, \end{array} \right. \label{eq:scal}
\eq
with a screened $\scalin\simeq1$ and, for $\gamma=1$,~\cite{terukina:13,lombriser:14a}
\bqa
 \scalout & = & \frac{8\pi G\rhos\rs^3}{3+2\omega}\ln\left(\frac{r+\rs}{\rc+\rs}\right)\frac{1}{r} \nonumber\\
 & & + (1-\scalenv)\frac{\rc}{r} + \scalenv, \\
 \rc & = & \frac{8\pi G\rhos\rs^3}{3+2\omega}\frac{1}{1-\scalenv}-\rs.
\eqa
Here, $\scal$ is the Jordan frame chameleon scalar field, $\scalenv$ is its value in the environment of the halo, characterizing an environmental dependence of the gravitational modification, and $\rc$ is the chameleon screening scale below which $\scal\simeq1$ and gravitational forces return to Newtonian.
Furthermore, we use a constant Brans-Dicke parameter $\omega$, the gravitational constant $G$, and set the speed of light in vacuum to unity.
The kinematically inferred effective gravitational coupling then becomes~\cite{schmidt:10,lombriser:14a}
\bq
 \Geff(r) = \left\{ 1 + \frac{\Theta(r-\rc)\epsilon}{3+2\omega}\left[ 1 - \frac{M(\rc)}{M(r)} \right] \right\} G, \label{eq:geff}
\eq
where $M(r)$ is the mass enclosed at radius $r$, obtained by integration of Eq.~(\ref{eq:density}).
$\Theta$ is the Heaviside step function and $\epsilon\in[0,1]$~\cite{hui:09} describes a possible self-screening effect of the object exposed to $\scal$ due to its own potential well.
Hereby, $\epsilon=0$ if the object is screened, $\epsilon=1$ if it is unscreened, and $\epsilon$ taking on a value in between in case of partial screening.
Note that Eq.~(\ref{eq:geff}) does not apply to scales beyond the Compton wavelength of the scalar field at the current cosmological background $\bscal_0$, $r\gtrsim\bar{m}_0^{-1}\sim\sqrt{(3+2\omega)(1-\bscal_0)}~{\rm Gpc}$, where $\Geff$ returns to its Newtonian value $G$.
The scales of interest for dwarf spheroidals, however, lie below $\bar{m}_0^{-1}$.

\psec{Cores from cusps}
Importantly, we observe that $\Geff$ increases with $\Delta r=r-\rc$ for $\rc<r\lesssim\bar{m}_0^{-1}$ before saturating at $\rc\ll r\lesssim\bar{m}_0^{-1}$ with $G_{\rm max}=(4+2\omega)/(3+2\omega)$ when $\epsilon=1$.
Hence, in the presence of a chameleon field, kinematic tracers which lie at different radii in this transition region, or in different regimes of constant $\Geff$, experience an additional dynamical contribution to the underlying matter distribution.
More specific, given a measurement of the luminosity-averaged velocity dispersions $\sigma_{vi}^2$ and the half-light radii $r_{{\rm h}i}$ of two distinct stellar subcomponents $(i=1,2)$ in a halo, the slope of the inferred dynamic mass profile may be characterized by
\bq
 \Gamma = \frac{\ln[M(r_{{\rm h}2})/M(r_{{\rm h}1})]}{\ln(r_{{\rm h}2}/r_{{\rm h}1})} + \frac{\ln[\Geff(r_{{\rm h}2})/\Geff(r_{{\rm h}1})]}{\ln(r_{{\rm h}2}/r_{{\rm h}1})}, \label{eq:gamma}
\eq
where we assume $\sigma_{vi}^2\simeq2\Geff(r_{{\rm h}i})M(r_{{\rm h}i})/(5r_{{\rm h}i})$~\cite{walker:11,schmidt:10,lombriser:12}.
The first term in Eq.~(\ref{eq:gamma}) corresponds to the slope inferred in Newtonian gravity whereas the second term only contributes if $\Geff$ differs between the two tracers.
In Newtonian gravity or the fully unscreened region, therefore, from Eq.~(\ref{eq:density}), $\Gamma\leq3-\gamma$.
Thus, for a cuspy NFW profile ($\gamma=1$) and core ($\gamma=0$), the slope is $\Gamma\leq2$ and $\Gamma\leq3$, respectively.
In this context, we identify three scenarios in which chameleon modifications may lead to an apparent core-like matter density profile ($\Gamma>2$) although $\gamma=1$:
\begin{itemize}
 \item[(i)] The inner and outer stellar populations lie in the transition region of the halo, where gravitational forces become increasingly enhanced with growing distance from the halo center.
 \item[(ii)] The inner stellar population lies in a screened region, whereas the outer population lies in an unscreened or partially unscreened region of the halo.
 \item[(iii)] Both the inner and outer stellar population lie in a unscreened region of the halo but the stars in the inner population are self-screened whereas the stars in the outer population are (partially) unscreened.
\end{itemize}

\psec{Fornax and Sculptor}
\begin{figure*}
 \centering
 \resizebox{\hsize}{!}{
  \resizebox{\hsize}{!}{\includegraphics{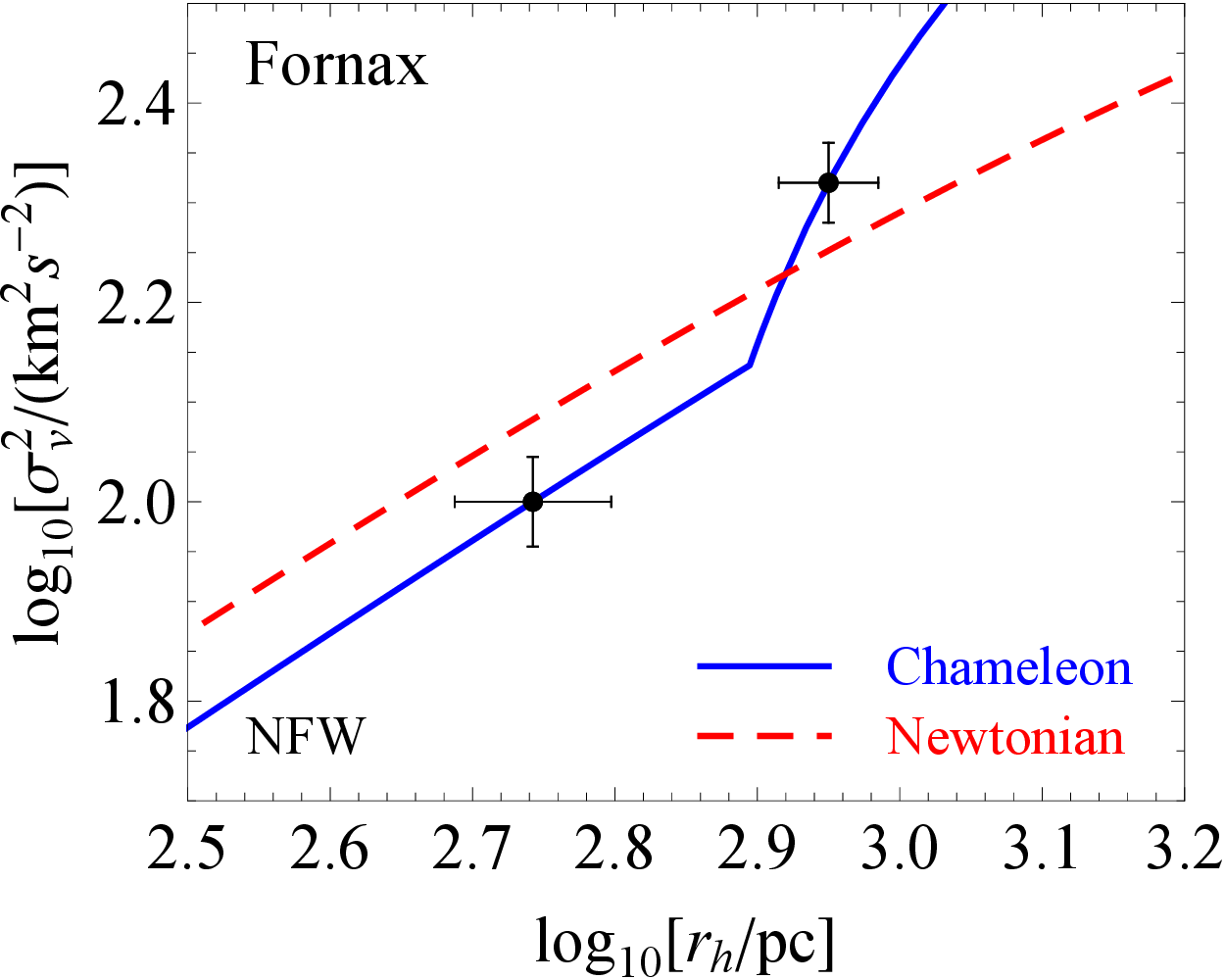}}
  \resizebox{\hsize}{!}{\includegraphics{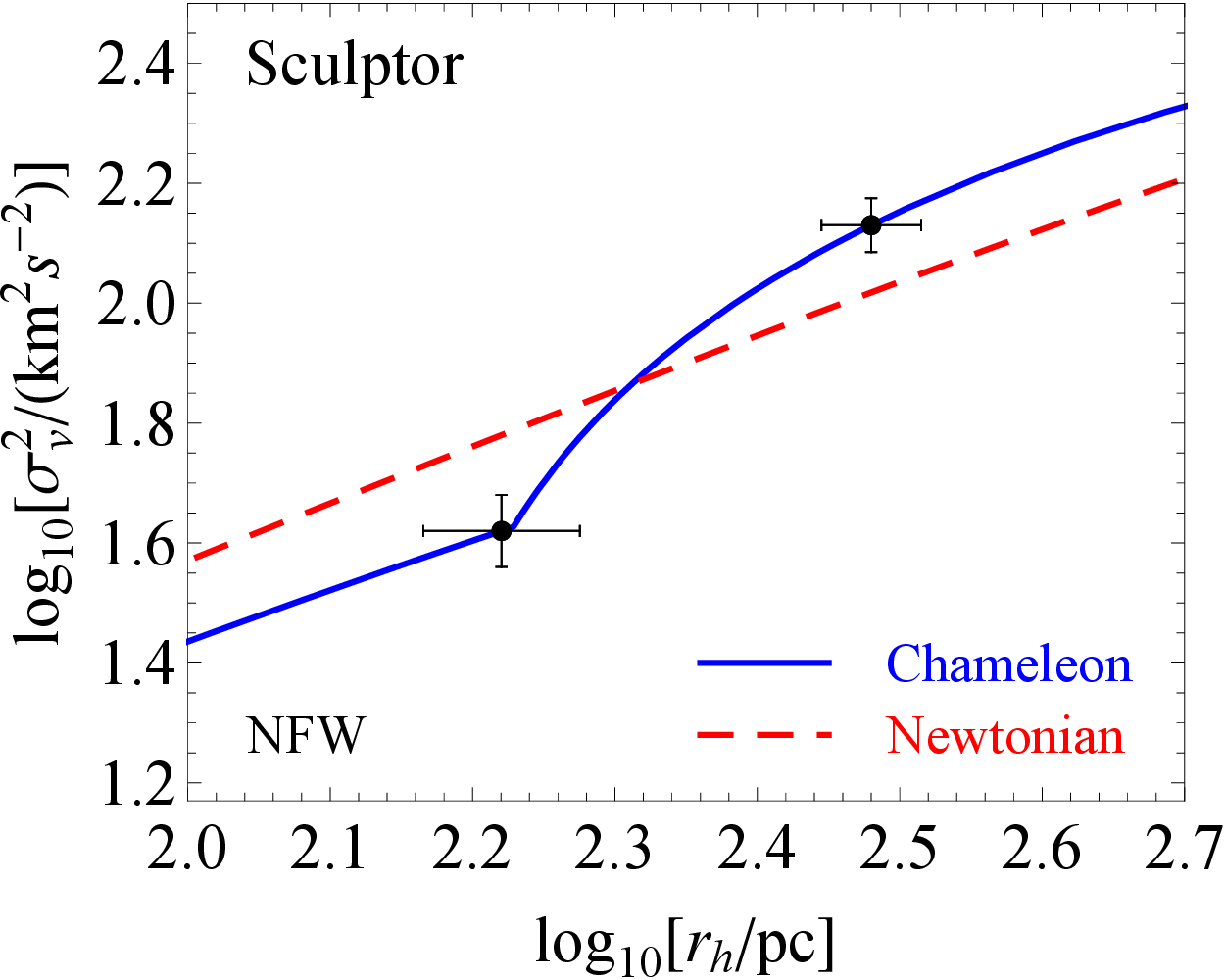}}
  \resizebox{0.997\hsize}{!}{\includegraphics{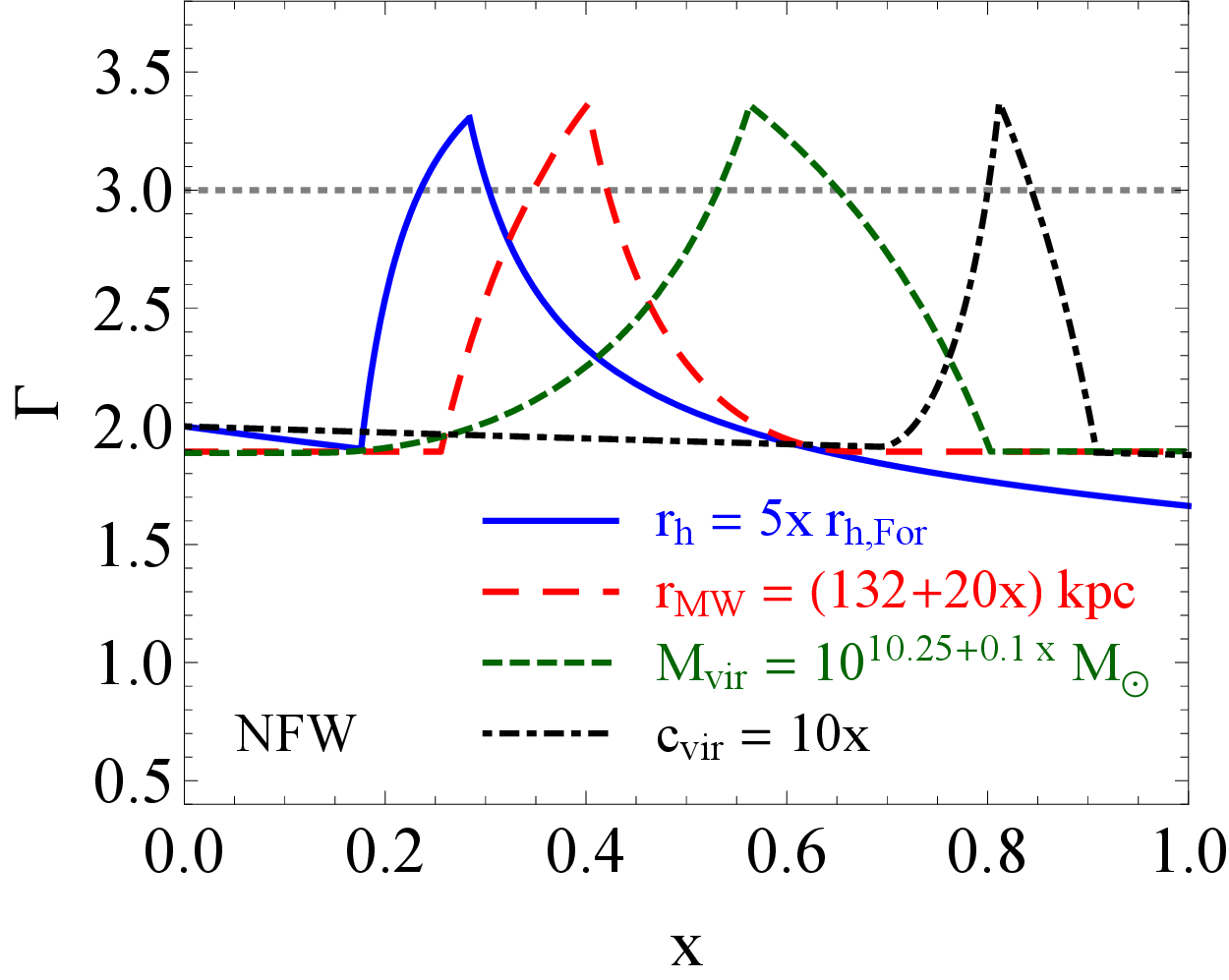}}
 }
 \caption{
  Velocity dispersions and half-light radii of the chemo-dynamically discriminated stellar subcomponents of Fornax (\emph{left panel}) and Sculptor (\emph{middle panel}) measured by Ref.~\cite{walker:11}.
  Whereas the assumption of a cuspy NFW matter density profile $(\gamma=0)$ does not provide good fits in Newtonian gravity, it becomes compatible with observations in chameleon gravity.
  \emph{Right panel:} The slope $\Gamma$ of the dynamic NFW mass profile of Fornax inferred from its two red giant populations when alternately varying its location in the Milky Way $r_{\rm MW}$, the position of the mean half-light radius $r_{\rm h}$ of the stars in the dwarf, as well as its virial mass and concentration while keeping other parameters fixed.
  In general, in both gravitational models $\Gamma$ becomes larger if baryonic feedback flattens the matter distribution $(0\leq\gamma<1)$.
  In Newtonian gravity $\Gamma>3$ would be interpreted as a depletion of matter in the interior of the dwarf.
  Hence, an observation of such an apparently pathological scenario may serve as a smoking gun for modified gravity.
  }
\label{fig:figure}
\end{figure*}
To demonstrate how chameleons can core dwarfs with cusps we consider the dark matter dominated Fornax and Sculptor dwarf spheroidals in the Milky Way.
Hereby, we use the measurements of $\sigma_{vi}^2$ and $r_{{\rm h}i}$ of two chemo-dynamically discriminated red giant populations in each of the two dwarfs (see Fig.~\ref{fig:figure}), which yield $\Gamma_{\rm For}=2.61^{+0.43(+1.07)}_{-0.37(-0.68)}$ and $\Gamma_{\rm Scl}=2.95^{+0.51(+1.22)}_{-0.39(-0.70)}$ at the 68\% (95\%) confidence level~\cite{walker:11}.
Note that these slopes are also robust under triaxiality of the cold dark matter halos~\cite{laporte:13}.

First, we provide a fit for Newtonian gravity assuming a cuspy NFW matter density profile.
We use free mass and concentration, which can be related to $\rhos$ and $\rs$, for each dwarf spheroidal.
Note, however, that there is a degeneracy between mass and concentration that yields $\Gamma\sim2$ with the appropriate ordinate in $\sigma_v^2$~\cite{penarrubia:07}.
We chose $M_{\rm For}=10^{10}~\Msun$, $c_{\rm For}=12.4$, $M_{\rm Scl}=10^{10}~\Msun$, and $c_{\rm Scl}=16.3$, where we set $\Deltavir=390$.
The corresponding stellar velocity dispersions for Fornax and Sculptor are shown in Fig.~\ref{fig:figure}.
Next, we refit the system assuming the presence of a chameleon field. 
To account for the environmental effects from the Milky Way halo on the gravitational modifications in the dwarfs we need to fit both galaxies simultaneously.
Hence, we trace the chameleon field from the environment of the Milky Way, the Local Group, to the positions of the Fornax and Sculptor dwarf spheroidals in the Milky Way halo, and to the positions of each of their two measured stellar subcomponents in the dwarfs.
The chameleon field in the entire system is then determined by $\scalenv=\scal_{\rm LG}$ and the Brans-Dicke constant $\omega$.
Note that $\scal_{\rm LG}$ can, in principle, be related to the scalar field value of the cosmological background $\bscal_0$ if tracing the field in the Local Group into its cosmological environment.
Importantly, both $\bscal_0$ and $\omega$ are universal parameters of the gravitational theory and once fixed they apply everywhere in the universe.
As we are not interested here in providing accurate measurements of these values, we will adhere to some simplification.
First, in order to estimate the scalar field profile in the Milky Way, we adopt a NFW profile.
Second, we set the virial mass and concentration of the Milky Way to $M_{\rm MW}=1.26\times10^{12}~\Msun$ and $c_{\rm MW}=9.5$~\cite{mcmillan:11}.
Note that these values have been inferred assuming Newtonian gravity and that the exact numbers are still debated.
The adoption of a Newtonian mass estimate is a conservative choice as accounting for enhanced forces reduces the inferred mass such that the halo becomes less screened with a transition more likely to occur.
Hence, a change of this mass estimate only causes a shift in $\scal_{\rm LG}$ and $\omega$ for the mechanism discussed here to become applicable.
Finally, we fix the positions $r_{\rm i}$ of Fornax and Sculptor at a distance of $r_{\rm For}=138~{\rm kpc}$ and $r_{\rm Scl}=79~{\rm kpc}$ from the Milky Way center, respectively~\cite{walker:11}.
The environmental scalar field values $\scalenv$ for the two dwarfs are then set by $\scal_{\rm MW}(r_{\rm i})$.
We adopt a partially screened charge $\epsilon=2/3$ for the red giant stars in the dwarfs.
This value approximately corresponds to a scenario in which the cores of the red giants are self-screened and their envelopes unscreened~\cite{jain:12}.
Note that the exact level of self-screening is not crucial for scenarios (i) and (ii) to work, as long as the stars are at least partially unscreened and experience some enhanced gravitational force.
Using Eqs.~(\ref{eq:scal}) through (\ref{eq:geff}), we determine the scalar fields $\scal_{\rm For}(r)$ and $\scal_{\rm Scl}(r)$ and the corresponding $\Geff(r)$ acting on the stellar subcomponents in Fornax and Sculptor, respectively.
In summary, in order to perform the chameleon fit we use the two additional, global parameters $\omega$ and $\scal_{\rm LG}$, together with %$\rhos$ and $\rs$
the mass and concentration of each dwarf spheroidal.

The chameleon modification allows a perfect fit
through the data points for $M_{\rm For}=2.1\times10^{10}~\Msun$, $M_{\rm Scl}=7.3\times10^{8}~\Msun$, $c_{\rm For}=8.8$, $c_{\rm Scl}=24.8$, $1-\scal_{\rm LG}=2.46\times10^{-6}$, and $\omega=-1.305$.
Note that these values satisfy the Solar System constraint $(1-\scal_{\rm LG})/10^{-6}\lesssim5/(6+4\omega)\approx6.4$~\cite{lombriser:13c} as the corresponding Milky Way chameleon screening scale lies at $\rc\simeq60~{\rm kpc}$, which is beyond the location of the Solar System $(\sim8~{\rm kpc})$.
Furthermore, the chameleon parameters found here are either compatible with or have not been constrained by cosmological~\cite{lombriser:14a} and astrophysical observations such as strong lensing~\cite{smith:09t}, rotation curves~\cite{vikram:14}, or standard candles~\cite{jain:12}.
Moreover, the scales of the systems considered here are well below the Compton wavelength of the scalar field $\bar{m}_0^{-1}\sim{\rm Mpc}$ for which Eq.~(\ref{eq:geff}) applies.
It is worth noting that in Fig.~\ref{fig:figure} both scenarios (i) and (ii) are approximately represented.
That is, (i) two stellar populations which reside in the transition region between screened and unscreened gravitational modifications in Sculptor, and (ii) a fully screened inner stellar population with a partially unscreened outer stellar population in Fornax.
Scenario (iii) is not represented, as the kinematic tracers are red giants for which we have approximated $\epsilon=2/3$ for a (partially) unscreened environment.
For different stars, self-screening may have to be described more accurately, which, however, requires more information on the properties of the observed stars and their environments.
Considerable work has been conducted in determining the self-screening effects of stars~\cite{chang:11,davis:11,jain:12,sakstein:13}, which may become important when analyzing the stellar dynamics for further dwarf spheroidals.

\psec{Smoking gun for modified gravity}
The dwarf spheroidals of the Milky Way are located at about $(20-600)~{\rm kpc}$ from the Galactic Center~\cite{mcconnachie:12}.
Additional observations of the different dynamic mass profiles of these dwarfs may generally be divided into objects that lie below its chameleon screening scale, $\rc\simeq60~{\rm kpc}$, and above it.
That is, dwarfs that are environmentally screened and do not experience modified gravitational forces and dwarfs which show a modified gravity signature if not sufficiently self-screened, respectively.
Note that while the additional scale dependence introduced by the chameleon force can mimic a core-like matter distribution ($2<\Gamma\leq3$) even if $\gamma=1$, there remains a degeneracy with the unknown level of baryonic feedback in these dwarfs, which can reduce $\gamma$ and yield $\Gamma>2$.
This has to be accounted for in both the Newtonian and chameleon scenarios.
Importantly, however, in chameleon gravity the increase in the effective gravitational coupling with increasing distance from the dwarf center can cause the slope to become $\Gamma>3$.
This is independent of the level of baryonic feedback allowed in chameleon gravity as for $0\leq\gamma<1$, $\Gamma$ generally becomes even larger, which can be seen from Eqs.~(\ref{eq:geff}) and (\ref{eq:gamma}).
We conservatively set $\gamma=1$ and illustrate this behavior in the right-hand panel of Fig.~\ref{fig:figure} by alternately varying the location of Fornax in the Milky Way halo, the position of the mean half-light radius $r_h=(r_{h1}+r_{h2})/2$ of its stellar subcomponents, as well as the mass and concentration of the dwarf while keeping other parameters fixed.
In Newtonian gravity, a measurement of such a slope would be interpreted as a pathological scenario of matter depletion in the dwarf center.
Hence, dynamical masses with $\Gamma>3$ may serve as a smoking gun for the presence of a chameleon field or a similar modification of gravity.
On the other hand, since dwarf spheroidals are located at larger distances from the the Galactic Center, a measurement of their dynamic mass profiles will allow an improvement over Solar System constraints.

\psec{Discussion}
Modifications of gravity on large scales, which potentially may provide an explanation for cosmic acceleration without invoking a cosmological constant, need to be shielded on local scales to satisfy Solar System tests.
This naturally introduces a transition between the unscreened and screened regimes, which implies an additional scale dependence in the dynamic mass profiles with respect to the underlying matter distributions of self-gravitating systems.
As an example of such a screening mechanism, we adopt here the chameleon model.
Tracing the chameleon field in the Milky Way, we simultaneously fit the velocity dispersions of chemo-dynamically distinct red giant populations in the dark matter dominated Fornax and Sculptor dwarf spheroidals under the assumption of a universal and cuspy NFW matter density profile.
Whereas for Newtonian gravity the NFW profiles do not provide good fits to the data, the observed core-like dynamic mass distributions pose no problem to chameleon models, which introduce only two new and universal parameters, the scalar field value in the cosmological background and the coupling of the scalar field to matter.
Moreover, chameleon models predict the existence of slopes between stellar subcomponents that in Newtonian gravity would be interpreted as a depletion of dark matter in the dwarf center.
If observed, this effect can serve as a smoking gun for modified gravity.
In general, measuring the dynamic mass profiles of the Milky Way dwarfs will improve constraints on chameleon gravity beyond those inferred from Solar System tests.

\begin{acknowledgments}
We thank Bhuvnesh Jain and Matthew Walker for useful discussions.
This work has been supported by the STFC Consolidated Grant for Astronomy and Astrophysics at the University of Edinburgh.
Please contact the authors for access to research materials.
\end{acknowledgments}

\vfill
\bibliographystyle{arxiv_physrev}
\bibliography{chameleoncorecusplib}

\end{document}